# SU(3) Lattice Gauge Theory With Adjoint Action At Nonzero Temperature[*]

T. Blum[a], Carleton DeTar[b], Urs M. Heller[c], Leo Kärkkäinen[a] and D. Toussaint[a]

[a]Department of Physics, University of Arizona, Tucson, AZ 85721, USA

[b]Department of Physics, University of Utah, Salt Lake City, UT 84112, USA

[c]SCRI, The Florida State University, Tallahassee, FL 32306-4052, USA

We study the thermal phase diagram of pure SU(3) gauge theory with fundamental and adjoint couplings. We improve previous estimates of the position of the bulk transition line and determine the thermal deconfinement transition lines for $N_t = 2, 4, 6$, and $8$. For $N_t > 4$ the deconfinement transition line splits cleanly away from the bulk transition line. With increasing $N_t$ the thermal deconfinement transition lines shift to increasingly weaker coupling, joining onto the bulk transition line at increasingly larger $\beta_a$ in a pattern consistent with the usual universality picture of lattice gauge theories.

## 1. Introduction

The phase diagram of fundamental–adjoint pure gauge systems,

$$S = \beta_f \sum_P [1 - \frac{1}{N} \text{ReTr} U_P] + \beta_a \sum_P [1 - \frac{1}{N^2} \text{Tr} U_P^\dagger \text{Tr} U_P], \qquad (1)$$

is considerably more complicated than the one with only a fundamental coupling. Early studies found that it has first order (bulk) transitions in the region of small $\beta_f$ [1,2]. In particular a line of first order transitions points towards the fundamental axis terminating for SU(3), as discussed in Sec. 2, at $(\beta_f, \beta_a) = (4.00(7), 2.06(8))$ and extending as a roughly straight line of bulk crossovers beyond the endpoint.

The non-trivial phase structure in the fundamental-adjoint plane, and in particular the critical endpoint, has been shown to be associated with, or even responsible for, the dip in the discrete $\beta$-function of the theory with standard Wilson action, which occurs in the region where the bulk crossover line crosses the fundamental axis. The bulk transition might also mask the thermal deconfinement transition, at least for lattices with small temporal extent $N_t$.

In a recent paper Gavai, Grady and Mathur showed that at nonzero temperature the deconfinement transition in pure gauge SU(2) is connected to the end point of the bulk transition line [3]. They were not able to distinguish the first order bulk line from the $N_t = 4$ and 6 thermal deconfinement transition lines in their simulations. Universality in the continuum limit requires that as $N_t$ is increased, the thermal phase transition line shift to weaker coupling, so that approaching the zero-temperature weak-coupling limit along any line in the fundamental-adjoint coupling plane leads to the same low temperature, confined theory. Thus the thermal phase transition lines could not remain anchored to a bulk transition line for higher $N_t$. More recently Mathur and Gavai [4] raised doubts about the very existence of a bulk phase transition in SU(2), characterizing even the first-order portion of the line as a thermal deconfinement transition that is displaced toward weak coupling with increasing $N_t$. Clarification is obviously needed.

Puzzled by the finding of Ref. [3], but also motivated by some unexpected results in the study of the high temperature behavior of full QCD with Wilson fermions [5] that might be explained by the fact that the fermion determinant induces, among other terms, an effective adjoint coupling, we decided to study pure SU(3) gauge theory with a fundamental-adjoint action at finite temperature in more detail [6]. We show that the first order bulk transition in pure gauge SU(3) in the fundamental-adjoint coupling plane indeed separates from the thermal deconfinement transition provided $N_t$ is made large enough.

---

[*]Talk presented by U. M. Heller at Lat94 Conference, September 27 – October 1, 1994, Bielefeld, Germany.

## 2. The phase diagram in the fundamental-adjoint coupling plane

We found it necessary to repeat the early calculation of Bhanot [2] to locate the endpoint of the bulk phase transition more accurately. This was done in a series of simulations with $8^4$ lattices by measuring the discontinuity in the average fundamental plaquette $E_f = \langle \text{Tr} U_P \rangle$ along the first order line. To support the identification as a bulk transition Fig. 1 shows $E_f$, obtained from various lattice sizes, as function of $\beta_f$ for $\beta_a = 2.25$. We note that $E_f$ jumps for all lattices at the same point.

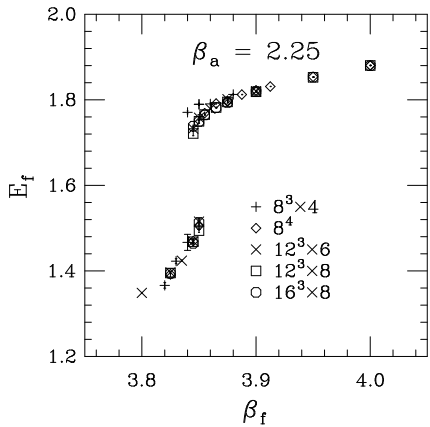

Figure 1. $E_f$ as function of $\beta_f$ for various lattice sizes, as indicated, at $\beta_a = 2.25$.

We obtained the gaps, shown in Fig. 2, from runs with hot and cold starts at the same value of the fundamental coupling $\beta_f$. The endpoint of the bulk transition line $(\beta_f^*, \beta_a^*)$ is then estimated by fitting to

$$\Delta E_f = c\, (\beta_a - \beta_a^*)^p. \qquad (2)$$

Such a fit works very well, as shown in Fig. 2, having $\chi^2 = 1.16$ for 2 degrees of freedom. In this region of the parameter space the bulk transition line is essentially straight, and from the endpoint value $\beta_a^* = 2.06(8)$ we infer the location of the endpoint of the bulk transition line as

$$(\beta_f^*, \beta_a^*) = (4.00(7), 2.06(8)), \qquad (3)$$

shown with our improved location of the bulk phase transition line in Fig. 3.

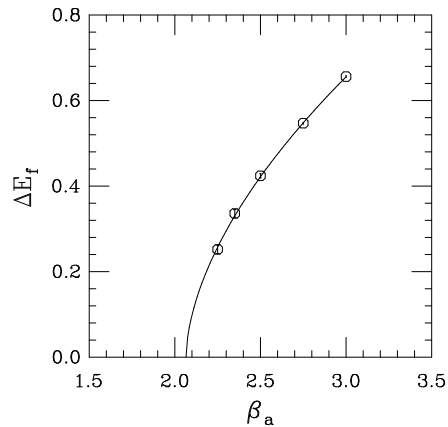

Figure 2. The average plaquette gap $\Delta E_f$ as function of $\beta_a$ together with the fit to determine the endpoint $\beta_a^*$.

## 3. Results at nonzero temperature

We studied the deconfinement transition on lattices with temporal extent $N_t = 2, 4, 6$ and 8, on lattices with aspect ratio $N_s/N_t = 2$ in the first 3 cases. All transition points we obtained are shown in Fig. 3. For $N_t = 2$ the $\beta_a = 0$ deconfinement transition continues smoothly into the bulk transition, which is shifted significantly from its large $N_t$ value, as $\beta_a$ increases. For $N_t = 4$ the deconfinement phase transition joins and is indistinguishable from the bulk transition, just as was found for SU(2) [3]. But for $N_t = 6$ at $\beta_a = 2.0$, there is a clear separation, with the bulk crossover occurring at $\beta_f = 4.035(5)$ and the thermal deconfinement transition at $\beta_f = 4.07(2)$. However, they appear to join at $\beta_a = 2.25, \beta_f = 3.850(5)$.

At larger $N_t$ the small size of the order parameter required additional care in locating the phase transition. We carried out a finite size analysis with simulations on $N_s^3 \times 8$ lattices with $N_s = 8, 12, 16$. In the confined phase $\langle|P|\rangle$ should vanish as $1/\sqrt{V_s}$ with increasing spatial volume, while in the deconfined phase it should attain a finite value. The finite size analysis is shown for $\beta_a = 2.25$ in Fig. 4. In this way we place the $N_t = 8$ deconfinement transition for $\beta_a = 2.0$ at $\beta_f = 4.135(15)$ and for $\beta_a = 2.25$ at $\beta_f = 3.89(1)$. This is still clearly separated from the bulk transition at $\beta_f = 3.850(5)$, whereas at $\beta_a = 2.50$, no clear separation is visible.

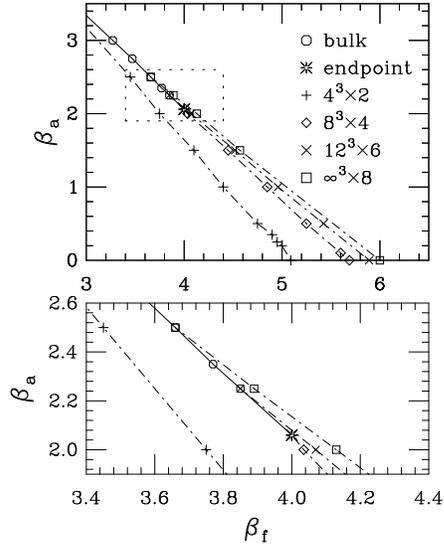

Figure 3. The phase diagram together with the thermal deconfinement transition points for $N_t = 2, 4, 6$ and $8$. The lower plot shows an enlargement of the region around the end point of the bulk transition.

We conclude that with increasing $N_t$ the deconfinement transition line is displaced toward weaker coupling, joining onto the bulk transition at larger and larger values of $\beta_a$, consistent with universality. However, trying to see low temperature continuum physics at larger values of the adjoint coupling $\beta_a$ requires larger lattices to avoid strong violations of asymptotic scaling associated with the bulk transition line.

### Acknowledgements

This work was partly supported by the DOE under grants No. DE-FG02-8SER40213, # DE-FG05-85ER250000 and # DE-FG05-92ER40742 and the U.S. National Science Foundation under grant PHY9309458. Computations were supported by SCRI, PSC and the Cornell Theory Center.

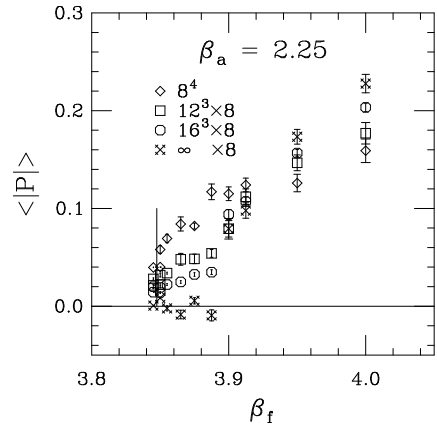

Figure 4. $\langle|P|\rangle$ for $N_t = 8$ and $N_s = 8, 12$ and $16$ and the extrapolation to $N_s = \infty$.